\documentstyle[aps,12pt]{revtex}
 \def\to{\rightarrow}
\def\gev{\mbox{GeV}}
\def\mev{\mbox{MeV}}
\def\ev{\mbox{eV}}

\def\EPJ{{\it Eur. Phys. J.} }

\def\NC{{\it Nuovo Cimento} }
\def\NP{{\it Nucl. Phys.} }
\def\PL{{\it Phys. Lett.} }
\def\PR{{\it Phys. Rev.} }
\def\PRL{{\it Phys. Rev. Lett.} }

\def\PTP{{\it Progr. Theor. Phys.} }

\def\ZP{{\it Z. Phys.} }

\def\frac#1#2{{\textstyle{{#1}\over {#2}}}}

\def\lsim{\mathrel{\rlap{\lower4pt\hbox{\hskip1pt$\sim$}}
    \raise1pt\hbox{$<$}}}
\def\gsim{\mathrel{\rlap{\lower4pt\hbox{\hskip1pt$\sim$}}
    \raise1pt\hbox{$>$}}}
\def\sqr#1#2{{\vcenter{\vbox{\hrule height.#2pt
         \hbox{\vrule width.#2pt height#1pt \kern#1pt
         \vrule width.#2pt}
         \hrule height.#2pt}}}}

\begin{document}
\preprint{IFT-P.0xx/2000\\ March 2000\\}
\title{Estimating $\sigma$-meson couplings from $D\rightarrow 3\pi$ decays}
\author{Claudio Dib $^{(1)}$ and Rogerio Rosenfeld $^{(2)}$}

\vskip 0.5cm
\address{(1) Universidad T\'ecnica Federico Santa Mar\'\i a\\
Departamento de F\'\i sica\\
1049-001 Valparaiso, Chile\\ E-mail address: cdib@fis.utfsm.cl}
\vskip 0.5cm
\address{(2) Instituto de F\'\i sica Te\'orica\\
R.\ Pamplona 145, 01405-900 S\~ao Paulo - SP, Brazil\\ E-mail
address: rosenfel@ift.unesp.br} \vskip 0.5cm
\date{\today}
\maketitle \vskip 1in

\begin{abstract}
Using recent experimental evidence from E791 on the sigma meson in
$D\to 3\pi$ decays, we study the relevant couplings in
$D\to\sigma\pi$ and $\sigma\to\pi\pi$ within the accepted
theoretical framework for non leptonic $D$ decays, finding an
overall consistency of the theory with the experimental data.
\end{abstract}
\newpage

\section{Introduction}
\label{int}
\parskip=7pt
\parindent=0pt

We know that QCD is the right theory for the strong interactions
at high energies \cite{QCD} and also it is well accepted as the
fundamental underlying theory at low energies. However, due to
infrared slavery, it is very difficult to apply this theory in the
hadronic realm because the resulting large coupling constant at
those lower energies makes it impossible to use perturbation
theory. Besides {\it ab initio} calculations, like lattice
techniques, one has to resort to an effective theory (chiral
perturbation theory, ChPT\cite{Gasser}) or sensible models
which describe the relevant degrees of freedom restricted by
the symmetries of the low energy domain.
An example of the latter is the sigma
model,\cite{Gell-Mann} which correctly describes
$\pi \pi$ scattering near threshold, to leading order in a
momentum expansion (and which coincides with ChPT at that order).
However, when one increases the center-of-mass energy,
the cross section starts to increase until resonances are reached.
There is a long standing controversy in the literature about the
existence of an isospin zero, broad scalar resonance in $\pi \pi$
scattering, the so-called sigma meson resonance \cite{reviews}.
Recently, a whole conference was devoted to the study of the
possible existence of the sigma resonance \cite{tornqvist}. The
controversy basically stems from the large width of the resonance,
which makes it difficult to discern if the shape of the spectrum
is actually due to a pole in the amplitude or to a result of other
effects in the $s$ and $t$ channels. Several experimental results
can be explained with the existence of such a
resonance\cite{experiment}. If it exists, it is a relevant degree
of freedom and must be incorporated into the analysis, together
with the conditions imposed by unitarity, chiral symmetry,
etc\cite{pipiscattering}. It also must show up in systems other
than the $\pi \pi$ system. For instance, it was pointed out ten
years ago that the sigma meson can have an important role in
explaining the $\Delta I = 1/2$ enhancement in $ K \to \pi \pi$
decays \cite{sanda}. Recently it has been found strong
experimental evidence that the sigma meson is very important in
the D-meson system, in the singly Cabibbo-suppressed  decay $D^+
\to \pi^+ \pi^+ \pi^-$, being responsible for approximately half
of the decays through the resonant sequence: $D^+ \to \pi^+\sigma
\to \pi^+ \pi^+ \pi^-$\cite{e791}. A best fit to the Dalitz plot
of this decay results in $m_\sigma = 483 \pm 31$ MeV and
$\Gamma_\sigma = 338 \pm 48$ MeV, where statistical and systematic
errors have been added in quadrature. We want to explore the
consequences of this experimental result in the context of the
well known theories of light mesons and weak decays. While we were
preparing this work, another study appeared on the effect of the
$\sigma$ in different weak processes\cite{Polosa}. They conclude
that the new result of E791, with a smaller value of the
non-resonant contribution to $\Gamma(D \to 3 \pi)$ is in better
agreement with the well measured $\Gamma(D \to 2 \pi)$, to which
they relate via PCAC. One should notice however, that PCAC must be
used with care, since it is valid only when the pion is soft, and
not in the entire kinematic range of these decays. Recently, the
$D \to \sigma \pi$ process was also studied in the context of a
constituent quark-meson model \cite{gatto}. Here we study the
specific consequences of a sigma particle and its couplings
related to the $D\to 3\pi$ process, invoking two formulations:
chiral symmetry as expressed in the linear sigma model to relate
the $\sigma\pi\pi$ coupling to other observables and the
QCD-inspired phenomenological model of Bauer, Stech
and Wirbel\cite{BSW} to treat the non-leptonic $D$ decay.

This paper is organized as follows. In section II we extract the
effective couplings involved in this $D$ decay from the
experimental data. In section III we explore the consistency of
this sigma meson within the models for spontaneous chiral symmetry
breaking of strong interactions. In section IV we address the
problem of estimating the weak process $D\to\sigma\pi$ in the
standard model of weak interactions and within the treatment of
Bauer, Stech and Wirbel for $D$ decays. Our conclusions are
presented in section V.

\section{The couplings in the decay $D\to \sigma\pi\to 3\pi$}
 From the experimental result that a fraction $f = (44 \pm 10)$\% of the
$D^+ \to \pi^+ \pi^+ \pi^-$ goes through the resonant
channel\cite{e791}, one can estimate the amplitude of the weak
process $D^+ \to \pi^+\sigma$. By resonant decay in this process
one understands $D\to\sigma\pi^+\to\pi^+\pi^-\pi^+$, {\it i.e.}
a $\pi^+\pi^-$ pair in the final state appears through the
formation of an intermediate $\sigma$ (sigma) resonance. The
coupling $D-\sigma-\pi$ in this resonant process is, in general, a
function of the $q^2$ of the virtual sigma ({\it i.e.} the
invariant mass of the $\pi^+\pi^-$ pair). This issue would not be
a problem if the sigma were narrow, because then the amplitude
would be strongly dominated at $q^2 = m_\sigma ^2$, fixing the
couplings to a constant value on the mass shell of the
intermediate particle. As a leading approximation, we assume that
the coupling is not a strongly varying function of $q^2$ and fix
it on the mass shell of the sigma. A more precise estimate would
require a model for the D decay into three pions via a sigma.
Assuming constant couplings $g_{D \sigma \pi}$ and $g_{ \sigma \pi
\pi }$ we have for the resonant three-body decay width:
\begin{eqnarray}
\Gamma (D^+ \to \sigma \pi^+ \to \pi^+\pi^- \pi^+) = \frac{1}{2}
\frac{1}{2 m_D}\; g_{D \sigma \pi}^2 \; &g_{ \sigma \pi \pi }^2&
\int_{4 m_\pi^2}^{(m_D-m_\pi)^2} \frac{dq^2}{2\pi}\; \;
\frac{1}{8\pi}\lambda^{1/2}
\left(1,\frac{q^2}{m_D^2},\frac{m_\pi^2}{m_D^2}\right)
\label{INTEGRAL}\\ \nonumber & &\times \;
\frac{1}{8\pi}\lambda^{1/2}
\left(1,\frac{m_\pi^2}{q^2},\frac{m_\pi^2}{q^2}\right)
\frac{1}{(q^2-m_\sigma^2)^2 + \Gamma_\sigma(q^2)^2 m_\sigma^2}
,\end{eqnarray}
where we use the standard notation $\lambda(a,b,c)\equiv
a^2+b^2+c^2-2ab-2ac-2bc$, and where each factor
$1/(8\pi)\times\lambda^{1/2}$ is the phase space integral of the
corresponding two-body decay subprocess. The factor $\frac{1}{2}$
in front is due to the symmetry of the two $\pi^+$ in the final
state, which the integral consider as distinguishable. Also
$\Gamma_\sigma(q^2)\equiv \Gamma_\sigma^0 \times\left(m_\sigma /
q\right)\left(p^\ast(q^2)/p^\ast(m_\sigma^2)\right)$ is the
co-moving resonance width; here $p^\ast(q^2) = \sqrt{q^2/4-
m_\pi^2}$. A value for the strong coupling $g_{ \sigma \pi \pi }$
can be obtained from the sigma width $\Gamma_\sigma ^0$ by
considering that the $\sigma$ meson decays 100\% into $\pi\pi$,
two thirds of the time into charged pions. Using the general
expression for the decay of a J=0 into two J=0 particles in terms
of the coupling constant \cite{rosner}, we have

\begin{equation}
\frac{2}{3}\Gamma_\sigma^0 = \Gamma(\sigma\to\pi^+\pi^-) = g_{
\sigma \pi \pi }^2 \; \frac{1}{8 \pi m_\sigma^2}\; p^\ast ,
\end{equation}
where $p^\ast=\frac{1}{2}\lambda^{1/2}\left( m_\sigma^2 , m_\pi^2,
m_\pi^2 \right) $ is the magnitude
of the 3-momentum of either of the final particles in the CM
frame. 

For our numerical results we will use the data $m_D = 1869$ MeV,
$m_\pi^+ = 139.6$ MeV\cite{PDG} and the experimentally extracted
values of the $\sigma$ resonance (mass and width) $m_\sigma = 483
\pm 31$ MeV and $\Gamma_\sigma^0 = (338 \pm 48 ) $ MeV\cite{e791}.
The value for the $g_{ \sigma \pi^+ \pi^- }$ coupling thus found
is:

\begin{equation}
g_{ \sigma \pi^+ \pi^- } = (2.59 \pm 0.19)~\gev.
\label{GSIGMAPIPI}
\end{equation}
This result, together with the numerical value of the integral
shown in Eq.~\ref{INTEGRAL}, which is $(2.55 \pm 0.55) \times
10^{-3} \; \gev^{-2}$ for all the appropriate values of the
parameters, allows us to get an estimate for the weak
$D-\sigma-\pi$ coupling:

\begin{equation}
g_{D \sigma \pi} =  654 \pm 120 \;\ev .\label{COUPLING}
\end{equation}
We have thus extracted this coupling using the same prescription
for the resonance that gave the experimental values of the sigma
mass and width cited above, and also assuming that the coupling is
independent of the virtuality of the intermediate $\sigma$ state.
The validity of our estimate is thus conditioned to the assumption
that the coupling for $q^2$ values away from the resonant peak
does not differ by much from its value at the peak and that the
shape of the resonance away from the peak is as prescribed in
Eq.~\ref{INTEGRAL}. Both approximations are clearly related and
constitute the old issue of the background underlying the
resonance.

To make a more crude estimate that does not deal with the
background issue, one could use a narrow width approximation to
estimate the coupling. Although this is {\it a priori} not a very
good approximation since the $\sigma$ resonance is not narrow, the
estimate is rather robust because it hides the width of the
resonance and only deals explicitly with its branching ratio.
Within this approximation the resonant $D$ decay width is:

\begin{equation}
\Gamma(D^+\to\pi^+\pi^+\pi^-)_{res}\approx
\frac{1}{2}\times\Gamma(D^+ \to \sigma \pi^+)\times
Br(\sigma\to\pi^+\pi^-),
\end{equation}
where again there is a symmetry factor $1/2$ due to the two
identical $\pi^+$. Now, $Br(\sigma\to\pi^+\pi^-)=2/3$ and
$\Gamma(D^+ \to \pi^+\pi^+ \pi^-)_{res}=f\times\Gamma(D^+ \to
\pi^+\pi^+ \pi^-)$, according to the experimental analysis. Using
the established data $Br(D^+ \to \pi^+\pi^- \pi^+) = (3.6 \pm 0.4)
\times 10^{-3}$ and $\tau_{D^+} = (1.06 \pm 0.02) \times
10^{-12}$s \cite{PDG}, one thus obtains the estimate for the decay
$D^+\to\sigma\pi^+$ in this approximation:

\begin{equation}
\Gamma(D^+ \to \sigma \pi^+) =3 \times  f \times \Gamma(D^+ \to
\pi^+\pi^- \pi^+) = (2.94 \pm 0.75) \times 10^{-12}\; \mev .
\end{equation}
    From this rate we can again extract the $g_{D\sigma\pi}$ coupling
using the general expression for the decay of a J=0 into two J=0
particles:

\begin{equation}
g_{D\sigma\pi} ^2 = \frac{8 \pi m_D^2}{p^\ast} \Gamma(D^+ \to
\sigma \pi^+), \label{AMPLITUDE}\end{equation}
where $p^\ast= 1/2\times\lambda^{1/2}\left(m_D ^2, m_\sigma^2 ,
m_\pi^2 \right)$ is again the 3-momentum of either of the final
particles in the CM frame. We thus obtain

\begin{equation}
g_{D\sigma\pi} \approx 548 \pm 70 \; \ev .\label{NARROWVALUE}
\end{equation}

One can notice that the central value of this estimate is only
16\% smaller than our more elaborate estimate of
Eq.~\ref{COUPLING}. In this sense, however crude the narrow width
approximation could be considered {\it a priori}, it is indeed
quite robust, since the result for $g_{D \sigma \pi}$ thus
obtained does not differ much from our more elaborate estimate, in
which the full resonance shape is taken into account.

\section{The $\sigma$ meson in the linear sigma model}

Having estimated the couplings directly from the
experimental data, let us see the consequences of these results
within the theory. First, consider the $\sigma \pi\pi$ coupling.
We are here in the low energy regime of strong interactions.
Consider then the linear sigma model as the framework, with
the $\sigma$ meson as the scalar ($J^{PC}=0^{++}$) that
remains from the breakdown of chiral symmetry. It is sufficient
to consider QCD with the lightest quarks $u$ and $d$ only, so that
the chiral symmetry in question is $SU(2)_L\times SU(2)_R$ broken
down to $SU(2)_V$ or isospin. The meson sector comprises the massive
$\sigma$ scalar and the three pions (pseudoscalars) $\pi^a$, $a=1,2,3$,
which play the role of Goldstone bosons. Using the notation of
ref.~\cite{Peskin} the quartic coupling of the potential,
$\lambda$, and the vacuum expectation value of the field, $v$,
determine all the physical parameters, like the mass of the sigma
as $m_{\sigma}^2 = 2\lambda v^2$,
the $\sigma \pi \pi$ coupling as $g_{\sigma\pi\pi} = 2\lambda v$ and
the (charged) pion decay constant as $f_\pi = \sqrt{2} v$.  The
linear sigma model thus predicts the $g_{\sigma\pi\pi}$ coupling
in terms of the mass of the $\sigma$ particle and the pion decay
constant:

\begin{equation}
g_{ \sigma \pi \pi } = \frac{\sqrt{2} m_\sigma^2}{f_\pi} = ( 2.54
\pm 0.01) \; \gev ,
\end{equation}
which is surprisingly close to the value deduced from the data
on resonant $D$ decay. Therefore, there is an apparent consistency
of the linear sigma model description for the $\pi\pi$ resonance in
$D$ decays. However, a word of caution is due here: the
validity of a perturbative calculation within a linear sigma model
treated as a quantum theory is questionable if the mass of the
scalar $\sigma$ is large ({\it i.e.} if $m_\sigma\sim f_\pi$),
which is precisely the case in the real world. A perturbative
calculation could be trusted if the expansion parameter is much
smaller than unity. In the sigma model, the
expansion parameter is $\lambda$, or more precisely
$6\lambda/(16\pi^2)$. From the data on $m_\sigma$ and $f_\pi$
it is easy to see that the value for the expansion parameter is:

\begin{equation}
\frac{6\lambda}{16\pi^2} \approx 0.5 ,
\end{equation}
which is uncomfortably close to unity. This is precisely the
reason why one would prefer to use a non-linear version of the
model as a perturbative theory, where the heavy sigma is integrated
out and only the light particles (the pions) are considered.
The effective interaction among the pions is then explicitly weak
at low momenta, vanishing at threshold (a feature which is evidence that
chiral symmetry is spontaneously broken at low energies and that
the pions are the corresponding Goldstone bosons). Here we used
the linear sigma model with the sole purpose of exhibiting the
sigma particle in the theory.

\section{The $D^+ \to \sigma \pi^+$ form factor in the BSW Model }
\label{model}
The coupling $g_{D\sigma\pi}$ found in Eq.~\ref{COUPLING}, which
has dimensions of mass, is an effective result due to an
underlying weak interaction process. In this section we want to
examine this coupling at that fundamental level. We therefore
estimate the relevant matrix elements of weak interactions that
contribute the $D^+ \to \sigma \pi^+$ decay. However, because of
the effect of strong interactions, the asymptotic states are not
the elementary quarks but composite hadrons, meaning that the
analysis cannot be done purely at the level of fundamental
interactions treated perturbatively. Only the weak interaction can
be considered at leading order, but to take into account strong
interactions we require of a sensible model. We thus study the
problem within the well accepted model of Bauer, Stech and Wirbel
(BSW) for non-leptonic D decays\cite{BSW}. For our purposes, the
$\sigma$ particle will be a $J^{PC}=0^{++}$ isoscalar with quark
content $(\bar u u + \bar d d)$. In order to tackle non-leptonic
decays, one has to use the operator product expansion to construct
an effective weak hamiltonian containing local four-quark
operators with Wilson coefficients $c_i(\mu)$ that can be computed
perturbatively at the appropriate energy scale $\mu$. In our case,
the relevant effective hamiltonian is:

\begin{equation}
H_w = \frac{G_F}{\sqrt{2}} V_{cd}^\ast V_{ud} \left\{ c_1(\mu)
(\bar d c)_{(V-A)} (\bar u d)_{(V-A)} + c_2(\mu) (\bar u
c)_{(V-A)} (\bar d d)_{(V-A)}
 \right\} ,
\end{equation}
where $(\bar q q')_{(V-A)} = \bar q \gamma_\mu \left( 1- \gamma_5
\right) q'$. In the BSW approach,the effective weak hamiltonian
density is built in terms of products of hadron currents which
mimic the underlying quark currents of the weak interactions:

\begin{equation}
H_{eff} = \frac{G_F}{\sqrt{2}} V_{cd}^\ast V_{ud} \left\{ a_1
J_\mu^{(\bar d c)} J^{\mu (\bar u d)} + a_2 J_\mu^{(\bar u c)}
J^{\mu (\bar d d)} \right\} .\label{HEFF}
\end{equation}
The coefficients $a_1$ and $a_2$ that accompany the products of
currents are now phenomenological constants of the model that must
be fitted from experiment. In this approach, the decay amplitude
factorizes into a product of two current matrix elements. We will
use this effective hamiltonian to study the $D^+ \to \sigma \pi^+$
decay process.

The factor proportional to $a_1$ in the matrix element
$\left<\sigma\pi^+|H_{eff}|D^+\right>$ involves two terms. One of
them is $\left< \sigma\pi^+|J_A ^{\mu(\bar u d)}|0\right>
\left<0|J_{A,\mu}^{(\bar d c)}|D^+\right>$, so-called
``annihilation'' term, which we neglect within the model as it is
prescribed, because it involves form factors of the light mesons
at high momentum ($q^2=m_D^2$)\cite{BSW}. We should emphasize here
that we are not neglecting the so-called {\it quark annihilation
diagrams} that play a role in explaining $D$ meson lifetime
differences and other controversial issues, but we are only
following a consistent prescription of the model. Indeed, in the
BSW model the coefficients $a_1$ and $a_2$ are only
phenomenological parameters and do not exactly correspond to the
Wilson coefficients $c_1(\mu)$ and $c_2(\mu)$ obtained in
perturbative QCD and associated with the quark amplitudes. The
other term, which is the only relevant term here, corresponds to
$\left< \sigma(p')|J_A ^{\mu (\bar d c)}|D^+(p)\right>
\left<\pi^+(q)|J_{A,\mu}^{\mu(\bar u d)}|0\right>$. After this
factorization, one needs the matrix element of the current (notice
that only the axial currents $J_A$ contribute, because of the
parity of the hadrons involved):

\begin{equation}
\left< \sigma(p')|J_A ^{\mu (\bar d c)}|D^+(p)\right>
 = F_1 ^{(D\sigma)}(q^2)
 \left( (p+p')^\mu - \frac{(m_D^2 -m_\sigma ^2)}{q^2}
 q^\mu \right)
+ F_0 ^{(D\sigma)}(q^2) \frac{(m_D^2 -m_\sigma ^2)}{q^2} q^\mu ,
\end{equation}
with $F_1(0) = F_0(0)$ in order to avoid a spurious singularity at
$q^2 = 0$. The two form factors $F_1$ and $F_0$ correspond to
transverse and longitudinal components of the current,
respectively. If these form factors are dominated by poles, those
of $F_1$ should be at axial vector meson masses and those of $F_0$
at pseudoscalar meson masses, all of $c\bar d$ flavor content.
However, only $F_0$ enters in our case, because
$\left<\pi^+(q)|J_A^\mu|0\right> = i f_\pi q^\mu$ is purely
longitudinal.

The factor of $a_2$ in the matrix element
$\left<\sigma\pi^+|H_{eff}|D^+\right>$ vanishes, because it
involves $\left<\sigma|J^\mu|0\right>$ which is identically zero
due to conservation of the vector current. Therefore, from all of
the above, the $D$-$\sigma$-$\pi$ coupling is:

\begin{equation}
\left<\sigma\pi^+|H_{eff}|D^+\right> \equiv g_{D\sigma\pi} =
\frac{G_F}{\sqrt{2}}
 V_{cd}^\ast V_{ud}\; a_1 \; F_0 ^{(D\sigma)}(m_\pi^2)
 \times(m_D^2 - m_\sigma ^2)\;  f_\pi .
\end{equation}
Using $V_{cd}^\ast V_{ud} = 0.21$, the model value $a_1 = 1.10 \pm
0.05$ fitted for D decays\cite{NS} and our value for the coupling
$g_{D\sigma\pi}$ in Eq.~\ref{COUPLING}, we get an estimate for the
axial form factor

\begin{equation}
F_0 ^{(D\sigma)}(m_\pi^2) = 0.79 \pm 0.15\; .
\end{equation}
We can safely extrapolate from $q^2 =m_\pi ^2$ to $q^2=0$, since
we do not expect poles at light masses for a charmed current:

\begin{equation}
F_0(0) = F_1(0) \approx 0.8 \pm 0.2\; ,
\end{equation}
which agrees within errors with the form factor of Ref.~\cite{BSW}
for $D\to\pi$ and is within the range of most of the other $D$
decays (0.6 to 0.8). A remarkable fact of this result is that the
form factors in Ref.~\cite{BSW} are calculated assuming the mesons
are bound states of the corresponding valence quarks, while in our
case one would hardly
 treat {\it a priori} such a short living resonance like the sigma
as a bound state, let alone using a wave function for it. As
expected, strong phases do not play a role in the $D^+ \to \sigma
\pi^+$ decay since the final state has only one definite value of
isospin , $I=1$. Strong phases in the $(3 \pi)$ final state have
already been taken into account by the introduction of resonances
in the fitting procedure to the Dalitz plot.

\section{Conclusion}
We have used the new experimental evidence for the $\sigma$ meson
in the non-leptonic $D^+\to\pi^+\pi^+\pi^-$ decay to estimate the
effective $g_{D \sigma \pi}$ and $g_{\sigma\pi\pi}$
couplings from the data and the resonance shape.
We then studied the consequences of the values of these couplings
within the accepted theoretical framework and found consistency
of the latter with the experimental data.

\vskip 0.3cm
{\bf Acknowledgments}
\vskip 0.3cm

\noindent We would like to
thank Ignacio Bediaga for the suggestion of and useful comments on
this work and to Jon Rosner for reading the manuscript. One of us
(R.R.) would like to thank the hospitality of the Departamento de
F\'\i sica of the Universidad Tecnica Federico Santa Maria,
Valparaiso, Chile, where this work was carried out. This work was
supported in part by CNPq, Pronex (Brazil) and FAPESP (S\~ao
Paulo) and by Fondecyt, Chile, grant No. 1980150.

\end{document}